\documentclass[a4paper,11pt]{article}
\pdfoutput=1 
\usepackage{jheppub} 

\usepackage[T1]{fontenc} 
\usepackage{caption} 
\usepackage{subcaption}

\usepackage[latin9]{inputenc}
\usepackage{float}
\usepackage{amsmath}
\usepackage{amssymb}
\usepackage{slashed}
\usepackage{graphicx}
\usepackage{esint}
 \usepackage{hyperref}
\usepackage{comment}
\usepackage{color}
\usepackage{microtype}
\usepackage{cleveref}
\usepackage{breakurl}
\usepackage{bbm}

\newcommand{\be}{\begin{equation}}
\newcommand{\ee}{\end{equation}}
\newcommand{\ben}{\begin{displaymath}}
\newcommand{\een}{\end{displaymath}}
\newcommand{\bea}{\begin{eqnarray}}
\newcommand{\eea}{\end{eqnarray}}
\def\K{K{\"a}hler}
   \newcommand{\rf}[1]{(\ref{#1})}
\newcommand{\vp}{\varphi}

\def\be{\begin{equation}}
\def\ee{\end{equation}}
\def\bea{\begin{eqnarray}}
\def\eea{\end{eqnarray}}
\def\ba{\begin{array}}
\def\ea{\end{array}}
\def\bit{\begin{itemize}}
\def\eit{\end{itemize}}

\def\La{\Lambda}

\def\vp{\varphi}

\newcommand{\g}{\gamma}
\renewcommand{\O}{\Omega}
\newcommand{\psibar}{\bar \psi}

\newcommand{\BS}{{\bf S}}
\newcommand{\BSbar}{{\bf \bar{S}}}
\newcommand{\BPhi}{{\bf \Phi}}
\newcommand{\BPhibar}{{\bf \bar \Phi}}
\newcommand{\BB}{{\mathbf{B}}}
\newcommand{\Op}{\Omega^\phi} 

\newcommand{\D}{\mathcal{D}}

 \makeatletter
\newcommand{\cG}{\mathcal{G}}

\def\rmi{{\rm i}}

\allowdisplaybreaks

\makeatother

\makeatletter
\DeclareRobustCommand{\rcite}[1]{%
  \rcite@aux#1,\@nil{#1}%
}
\def\rcite@aux#1,#2\@nil#3{%
  \if\relax#2\relax
    Ref.~\cite{#3}%
  \else
    Refs.~\cite{#3}%
  \fi
}
\makeatother

\hypersetup{
    colorlinks = true,
    citecolor = {blue},
    linkcolor = {blue},
    urlcolor = {blue},
}

 \title{\rm {\bf \huge  \boldmath Simple Sinflaton-less $\alpha$-attractors}}

\author{Renata Kallosh}
\author{and Yusuke Yamada}
\affiliation{Stanford Institute for Theoretical Physics and Department of Physics, Stanford University, Stanford, CA 94305, USA}
\emailAdd{kallosh@stanford.edu}
  \emailAdd{yusukeyy@stanford.edu}

\notoc

\abstract{We construct the simplest inflationary $\alpha$-attractor models in supergravity: it has only one scalar, the inflaton. There is no sinflaton since
the inflaton belongs to  an  orthogonal nilpotent superfield where  the sinflaton  depends on  fermion bilinears. When the local supersymmetry is  gauge-fixed,   these models have  only one single real scalar (the inflaton), a graviton and a massive gravitino. The sinflaton, sgoldstino  and inflatino  are all absent from the physical spectrum in the unitary gauge.
 The orthogonality condition leads to the simplest K\"ahler potential for the  inflaton, while preserving the Poincar\'e disk geometry of $\alpha$-attractors. The models are particularly simple in the framework of the $\overline {D3}$ induced geometric inflation.}

\begin{document}

\maketitle

 \newpage 
  \tableofcontents{}

 \parskip 6pt 
\section{Introduction}
Non-linearly realized supersymmetry of the Volkov-Akulov type  \cite{Volkov:1972jx}
plays an important role in  cosmological model building with de Sitter vacua/dark energy and near de Sitter inflationary evolution of the universe. A closely related to Volkov-Akulov supersymmetry is an approach based on a nilpotent superfield \cite{Rocek:1978nb,Ivanov:1978mx,Lindstrom:1979kq,Samuel:1982uh,Casalbuoni:1988xh}. Various more general constrained superfields were studied     in  \cite{Brignole:1997pe,Komargodski:2009rz,Kuzenko:2010ef}. 

A significant interest to these formal developments in non-linearly realized supersymmetry  started in applications to cosmology  in \cite{Antoniadis:2014oya,Ferrara:2014kva,DallAgata:2014qsj} when it became clear that certain difficulties in construction cosmological models with supersymmetric embedding can be solved with the help of  constrained superfields.  The new supergravity actions were constructed 
 which involve a supergravity multiplet coupled to a nilpotent chiral multiplet $\BS^2=0$. They  have natural de Sitter vacua \cite{Bergshoeff:2015tra,Hasegawa:2015bza}. In string theory the role of the $\overline {D3}$ brane is analogous to that of the nilpotent chiral multiplet $\BS^2=0$ in supergravity \cite{Kallosh:2014wsa}.

Many new and interesting properties of supersymmetric theories with constrained multiplets were studied during the last few years, see for example
\cite{Ferrara:2015tyn,DallAgata:2015zxp,Ferrara:2016een,Kallosh:2016hcm,DallAgata:2016syy,Cribiori:2019cgz}.
For our work here we will focus on the case when there is one nilpotent chiral superfield $\BS$
\be\BS^2=0
\label{nilp}\ee 
and one orthogonal  superfield $\BPhi$, which has a nilpotency degree 3,
\be
\BS (\BPhi- \BPhibar)=0\, ,\qquad (\BPhi- \BPhibar)^k=0, \qquad k\geq 3.
\label{ort}\ee
The orthogonal multiplet was first introduced in \cite{Brignole:1997pe} and studied in global supersymmetry in \cite{Komargodski:2009rz} and in local supersymmetry in \cite{Ferrara:2015tyn,Ferrara:2016een,Kallosh:2016hcm}. It was employed in the cosmological setting in \cite{Ferrara:2015tyn,Carrasco:2015iij} where it was recognized that the corresponding cosmological models are very simple, the inflatino and the sinflaton both absent. A string theory origin of constrained multiplets in the context of a
non-linearly realized spontaneously broken supersymmetry of the $\overline{D3}$-brane action in type IIB string theory was investigated in 
\cite{Kallosh:2016aep,Vercnocke:2016fbt}.

At the time when the cosmological models in \cite{Ferrara:2015tyn,Carrasco:2015iij} were studied, the most advanced  form of the $\overline{D3}$ induced geometric inflation~\cite{McDonough:2016der,Kallosh:2017wnt} was not known. It was therefore necessary in the description of the cosmological models to provide two holomorphic functions in the superpotential, $W= g(\Phi) + f(\Phi) S$ with the potential of the form $V= f^2(\vp)- 3g^2(\vp) $. The advantage of the geometric inflation ~\cite{McDonough:2016der,Kallosh:2017wnt} is that the basic information about the model is codified in the  \K\, geometry of the nilpotent superfield $\BS$, in $K_{S\bar S}(\Phi, \bar \Phi)$ as follows.
\be
K_{S\bar S}(\Phi, \bar \Phi) = {W_0^2 \over   3 W_0^2 + V(\Phi, \bar \Phi) }
\ee
and the superpotential can be a constant. These models are significantly simpler than the ones in \cite{Ferrara:2015tyn,Carrasco:2015iij} and lead directly to desired potentials.

However, it was was pointed out in \cite{Hasegawa:2017hgd,Hasegawa:2017nks}  that there might be a physics issue with the  cosmological models using constrained multiplets. It was at the time when there was a hope that LHC might discover supersymmetric particles and one would expect that the gravitino is light. In such case it was pointed out in 
\cite{Hasegawa:2017hgd,Hasegawa:2017nks} that the models with orthogonal nilpotent multiplet in \cite{Ferrara:2015tyn,Carrasco:2015iij}  as well the geometric models in \cite{McDonough:2016der,Kallosh:2017wnt}
 might  suffer from  gravitino overproduction after inflation.  At present   there is no compelling reason to assume that gravitino is light, since the superpartners of standard model particles have not been discovered so far.  Therefore, assuming that gravitino is not light, we may safely study the cosmological models with constrained superfields, as well as the geometric models of inflation with the purpose to establish a simple version of these supersymmetric cosmological models. 
 
Recent cosmological observations \cite{Akrami:2018odb} suggest  that the $\alpha$-attractor models of inflation \cite{Kallosh:2013yoa,Kallosh:2017wnt} remain at the sweet spot of the data. Therefore the purpose of this note is to present the simplest possible supergravity version of $\alpha$-attractor models of inflation. For this purpose we will combine the advantages of the orthogonal nilpotent superfield with the geometric model of inflation. In particular, we will define our new supergravity theories with one nilpotent multiplet $\BS^2=0$ and an inflaton orthogonal multiplet with $\BS (\BPhi-\BPhibar)=0$. We will present for these models   the \K\,function $\cG$, and also the \K\,potential $K$ and the superpotential $W$, in agreement with the definition of these functions
\be
\cG \equiv K + \ln W +\ln \overline W\, , 
\label{GKW}\ee
and 
\be
 V = e^ {\cal   G}  (\cG^{\alpha \overline \beta }  {\cal   G} _\alpha  {\cal   G} _{\overline \beta}  - 3 )  = e^K (K^{\alpha \overline \beta }  {D} _\alpha W  D_{\overline \beta}\overline{W}  - 3 W \overline W).
\ee

\section{Orthogonal  nilpotent superfield in supergravity}

We describe the constrained chiral superfields in supergravity following \cite{Ferrara:2016een}, \cite{Kallosh:2016hcm}, and we also use the notations of these papers.

Consider two unconstrained chiral multiplets, ${\bf S}$  and $\bf \Phi$  which have the following independent components each:  a holomorphic scalar, a left-handed spinor and an auxiliary field:
\be
{\BS}  : \qquad   \left\{ s_r+   {\rm i}  s_i, \, P_L \Omega^s,\, F^s \right\},
\ee
 \be
{\BPhi}  : \qquad   \left\{ \vp+{\rm i}b,\, P_L \Omega^\phi,\, F^\phi \right\}.
\ee
The first superfield $\BS$ in the context of our cosmological models is often called a stabilizer superfield. Its first component $s=s_r+  {\rm i}  s_i$ is associated with sgoldstino, the second one, a spinor $\Omega^s$ is a goldstino, and $F^s$ is an auxiliary field of the goldstino multiplet. 

The second multiplet $\BPhi$  in the context of our cosmological models is an inflaton multiplet. The first component, $\vp+{\rm i}b$ has an inflaton $\vp$ and a sinflaton $b$, the second component is an inflatino, a spinor $ \Omega^\phi$,  and finally $F^\phi$ is an auxiliary field of the inflaton multiplet.

We impose the conditions that $\BS$ is a nilpotent chiral superfield defined in eq.~\rf{nilp} and that $\BPhi$ is an orthogonal  real nilpotent superfield defined in eq.~\rf{ort}.

Using the supergravity multiplet calculus defined for the constrained superfields in~\cite{Ferrara:2016een} one finds from $\BS^2=0$ that  the scalar component of the nilpotent chiral multiplet $\BS$  is a bilinear combination of the fermions in the same multiplet, under condition that the auxiliary field  $F^s$ is not vanishing
\be \label{sgold}
s\equiv  s_r+  {\rm i}  s_i= {\overline \Omega^s P_L \Omega^s\over 2 F^s}, \qquad F^s\neq 0.
\ee
Orthogonality condition $ \BS \BB=0$, where $\BB=\frac{1}{2\rm i}(\BPhi-\overline{\BPhi})$, can be  resolved and one finds as shown in \cite{Ferrara:2016een}, \cite{Kallosh:2016hcm}  that {\it all component fields of the inflaton multiplet $b, \Omega^\phi, F^\phi$ with the exception of the inflaton field $\vp$} are not independent anymore. Namely, $b, \Omega^\phi, F^\phi$ depend on $\Omega^s$ and on $F^s$ and on the inflaton field $\vp$. We present explicit  solutions for $b, \Omega^\phi, F^\phi$, as functions of  
$\Omega^s$, $F^s$ and  $\vp$, in Appendix~\ref{appB}.

\section{Locally supersymmetric action}\label{LSA}
To present a complete locally supersymmetric action for a given \K\,function $\cG (\BS, \BPhi, \BSbar, \BPhibar)$,  or a \K\,potential  $K(\BS, \BPhi, \BSbar, \BPhibar)$ and a superpotential $W(\BS, \BPhi)$ where the superfields are constrained, one has to use the 
 the rules in \cite{Bergshoeff:2015tra,Hasegawa:2015bza,Kallosh:2015sea,Kallosh:2015tea,Schillo:2015ssx,Ferrara:2015tyn,Freedman:2016qnq,Freedman:2017obq}.
 For example the kinetic terms require to differentiate the  \K\,function $\cG (\BS, \BPhi, \BSbar, \BPhibar)$ before any constraints on the superfields are imposed,  to construct the complete locally supersymmetric action  for a given \K\,potential  and a superpotential. 
Once the full action is established, 
and  all kinetic terms and the potential are available, one can impose the constraints 
$\BS^2=0\, ,  \BS \BB=0
$ and their consequences which show that the superfields' components are not independent.

There are two possibilities to proceed in the context of the cosmological applications:

\begin{enumerate}
  \item One can focus {\it only on the bosonic action} of a locally supersymmetric  supergravity action and neglect all terms with fermions. 
  \be
\Omega^s=   \Op = \psi_\mu=0.
  \ee
  In such case, we can deduce already from eqs.  \rf{sgold},\rf{b},  \rf{inflatino},  \rf{Fp} that the bosonic action depends only on gravity and on the inflaton field. There is no sgoldstino scalar~$s$ and no sinflaton scalar~$b$ anymore since they are functionals of fermions,
 \be
\Omega^s=   \Op = \psi_\mu=0\, \qquad \Rightarrow \qquad s=b=0.
  \ee   
  \item  One can use the local supersymmetry  and  {\it gauge fix the action in the unitary gauge} \cite{Kallosh:2015sea,Ferrara:2015tyn} where gravitino is massive and sgoldstino is absent
\be
\Omega^s=0.
\ee
It follows\footnote{When the inflaton is not in an orthogonal multiplet, the gauge-fixing  $\Omega^s=0$ does not imply $\Op=0$, so that gravitino and inflatino actions are complicated, in general. In case the inflaton is in an orthogonal multiplet the inflatino vanishes in the gauge $\Omega^s=0$ due to eq. \rf{inflatino}.}
that in this gauge  that the sgoldstino $s$, the sinflaton $b$, the inflatino $\Op$ and the auxiliary field ${F}^\phi$ all vanish,
\be
\Omega^s=0,\, \qquad \Rightarrow \qquad s= b= \Op= {F}^\phi=0
\label{vanish}\ee
as one can see  from equations   \rf{sgold},\rf{b},  \rf{inflatino},  \rf{Fp}.
In this gauge therefore
\be
{\BS}|_{\rm unitary}  : \qquad   \left\{0, \, 0,\, F^s \right\},
\ee
 \be
{\BPhi}|_{\rm unitary}   : \qquad   \left\{ \vp,\, 0,\, 0 \right\},
\ee
i. e.
the nilpotent multiplet $\BS$ has only a non-vanishing auxiliary field whereas the inflaton multiplet $\BPhi$ has only a non-vanishing real scalar, an inflaton field.
\end{enumerate}
This is the reason why we find that the cosmological $\alpha$-attractor supergravity models with the orthogonal inflaton multiplet are  particularly simple. 
Note, however, that we cannot use the constraints on superfields in the  \K\,function $\cG (\BS, \BPhi, \BSbar, \BPhibar)$,  or in a \K\,potential  $K(\BS, \BPhi, \BSbar, \BPhibar)$ or in a superpotential $W(\BS, \BPhi)$. We can use the constraints in the action, where we encounter also the derivatives of these functions over the superfields, for example in kinetic terms for superfields and in the potential.

\section{Simple T-models with one nilpotent and one orthogonal  multiplet}\label{Tmodel}
Below we will mostly skip  the boldface notation,  we will use only for constraints on superfields. We impose the following constraints on T-model superfields

\be
\BS^2=0\, , \qquad \BS ({\bf Z}- {\bf \bar Z})=0\, , \qquad  \Rightarrow  \qquad  ({\bf Z}- {\bf \bar Z})^k=0 \, ,\quad k\geq 3.
\label{nilportT}\ee

\noindent The models are defined by a \K\, function $\cG$  so that the potential is
\be
\cG=- {1 \over 2} \cG_{Z\bar Z} (Z-\bar Z)^2  +  \cG_{S\bar S} \, S \overline S  +S + \overline S +\ln W_0^2\,  .
\label{G}\ee
Here  we take the \K\,metric as 
\begin{equation}
\cG_{Z\bar{Z}}=\frac{3\alpha}{(1-Z\bar{Z})^2}, \qquad \cG_{S\bar{S}} = {W_0^2 \over |F_S|^2+ f(Z\overline Z)}
\label{Metric}\end{equation}
According to Proposal 1.  in Sec.~\ref{LSA}, the scalar dependent part of the supergravity action with local supersymmetry depends on one real scalar $Z=\bar Z$ since the sinflaton field $b= {1\over 2\rm i} (Z-\bar Z) $ is a combination of fermions, as shown in eq. \rf{b}.
\be
{\cal L}_{\rm sc}= \frac{3\alpha}{(1-Z^2)^2}\partial Z \partial Z - V(Z).
\ee
Thus, there is no need here to stabilize the sinflaton scalar, since $(Z-\bar Z) $ is a product of fermions and drops from the bosonic part of the action. Here the potential  for the geometric field  is
\be
 V(Z)  =\Lambda   + f(Z^2),
\label{V}\ee
where 
\be
\Lambda \equiv |F_S|^2- 3W_0^2
\ee
is a cosmological constant.
The same action in terms of a  canonical field $\varphi$, where ${1\over 2} (Z+\bar Z) =\tanh\left(\frac{1}{\sqrt{6\alpha}}\varphi \right)$   is
\be
{\cal L}_{\rm sc}= \frac{1}{2} (\partial \vp)^2  -\left(\La   + f\left(\tanh^2 {\varphi\over \sqrt{6\alpha}}\right)\right ).
\label{sc}\ee
Now, according to  proposal 2 in Sec.~\ref{LSA}, we perform  a gauge-fixing of the total action. In such case we find that since many fields vanish according to \rf{vanish}, the total gauge-fixed action consists of a gravity part, gravitino part and a real scalar part
\be
{\cal L}_{\rm total}^{\text{gauge-fixed}}= {\cal L}_{\rm gravity}  + {\cal L}_{\rm gravitino} + {\cal L}_{\rm sc}.
\ee
In particular, the fermion spectrum has only a massive gravitino and no other fermions, like inflatino or goldstino, and there is only one real scalar besides the graviton in the bosonic sector.

We can take a simple function in the \K\,metric in \rf{Metric}
\be
f(Z\overline Z)= m^2 Z\bar Z
\ee
which results in a simplest T-model potential
\be
 V_T  =\La  + m^2 Z\overline Z= \Lambda + m^2 \tanh^2 {\varphi\over \sqrt{6\alpha}}.
\ee

\subsection{\K\, potential and  superpotential}
Instead of a \K\, function $\cG$ as shown in eqs. \rf{G},\rf{Metric} we can use a \K\, potential and a superpotential for presenting the models above, in agreement with eq. \rf{GKW}, 
\be
K= -{1 \over 2} K_{Z\bar Z} (Z-\bar Z)^2  +  K_{S\bar S} \, S \overline S  +(S + \overline S)  ,
\label{K}\ee
where
\begin{equation}
K_{Z\bar{Z}}=\frac{3\alpha}{(1-Z\bar{Z})^2}, \qquad K_{S\bar{S}} = {W_0^2 \over |F_S|^2+ f(Z\overline Z)},
\label{KZ}\end{equation}
and 
\be
W=W_0.
\ee
The total action in the unitary gauge $\Omega^s=0$ is very simple
\begin{align}
 e^{-1}{\cal L} & = \frac{1}{2\kappa^2} \left[ R(\omega (e )) -\psibar_\mu \gamma ^{\mu \nu \rho } D_\nu \psi _\rho
 +{\cal L}_{\rm SG,torsion} \right] \nonumber\\
 \cr
&\quad ~ +\frac{W_0}{2}  \psibar_{\mu  } \gamma ^{\mu \nu }\psi _{\nu  } - {1\over 2} (\partial \vp)^2-\left(\La   + f\left(\tanh^2 {\varphi\over \sqrt{6\alpha}}\right)\right),
\label{actionUG}
\end{align}
since
\be
e^K \Big |_{\BS^2=0, \, \BS ({\bf Z}- \overline{\bf Z})=0} =1\, ,  \qquad K_Z\Big |_{\BS^2=0, \, \BS ({\bf Z}- \overline{\bf Z})=0}=0.
\ee

\section{Simple E-models with one nilpotent and one orthogonal multiplet}\label{Emodel}
We impose the following constraints on E-model superfields

\be
\BS^2=0\, , \qquad \BS ({\bf T}- \overline{\bf T})=0\, , \qquad  \Rightarrow  \qquad  ({\bf T}- \overline{ \bf T})^k=0 \, ,\quad k\geq 3.
\label{nilportT}\ee

\noindent \noindent The E-models are defined by the following  \K\, function $\cG$ 
\be
\cG= -{1 \over 2} \cG_{T\bar T} (T-\bar T)^2  +  \cG_{S\bar S} S \overline S  +S + \overline S +\ln W_0^2\,  .
\label{GE}\ee
Here the \K\, metric is
\begin{equation}
\cG_{T\bar{T}}=\frac{3\alpha}{(T+\bar T)^2}, \qquad \cG_{S\bar{S}} = {W_0^2 \over |F_S|^2+ f\Big (1- {T+\bar T\over 2}\Big )}
\label{MetricE}\end{equation}
and the potential  for the geometric field  is
\be
 V  =\La   + f\Big (1- {T+\bar T\over 2}\Big ).
\label{VE}\ee
The potential for the canonical field $\varphi$, where ${1\over 2} (T+\bar T) =e^{- \sqrt { \frac{2}{{3\alpha}}}\varphi}$  is
\be
 V  =\La  + f\Big (1- e^{- \sqrt { \frac{2}{{3\alpha}}}\varphi}\Big ).
\ee

In particular we can take a simple function
\be
f\Big (1- {T+\bar T\over 2}\Big )= m^2 \Big (1- {T+\bar T\over 2}\Big )^2\ee
which results in a simplest E-model potential
\be
 V_E  =\La  + m^2 \Big (1- {T+\bar T\over 2}\Big )^2= \Lambda + m^2 \Big (1- e^{- \sqrt { \frac{2}{{3\alpha}}}\varphi}\Big )^2.
 \ee
 
 \subsection{\K\, potential and  superpotential}
Instead of a \K\, function $\cG$ as shown in eq. \rf{G} we can use a \K\, potential and a superpotential for presenting the models above, in agreement with eq. \rf{GKW}. 

\be
K= -{1 \over 2} K_{T\bar T} (T-\bar T)^2 
 +  K_{S\bar S}\,  S \overline S  +S + \overline S  ,
\label{K1}\ee
where
\begin{equation}
K_{T\bar T}=\frac{3\alpha}{(T+\bar T)^2}, \qquad K_{S\bar{S}} = {W_0^2 \over |F_S|^2+ f\Big (1- {T+\bar T\over 2}\Big )},
\label{KT}\end{equation}
\be
W=W_0.
\ee
The total action in the unitary gauge $\Omega^s=0$ is very simple,
\begin{align}
 e^{-1}{\cal L} & = \frac{1}{2\kappa^2} \left[ R(\omega (e )) -\psibar_\mu \gamma ^{\mu \nu \rho } D_\nu \psi _\rho
 +{\cal L}_{\rm SG,torsion} \right] \nonumber\\
 \cr
&\quad ~ +\frac{W_0}{2}  \psibar_{\mu  } \gamma ^{\mu \nu }\psi _{\nu  } - {1\over 2} (\partial \vp)^2-\Big (\La   + f\Big (1- e^{- \sqrt { \frac{2}{{3\alpha}}}\varphi}\Big )\Big ),
\label{actionUG}
\end{align}
since
\be
e^K \Big |_{\BS^2=0, \, \BS ({\bf T}- \overline{\bf T})=0} =1\, ,  \qquad K_T\Big |_{\BS^2=0, \, \BS ({\bf T}- \overline{\bf T})=0}=0.
\ee

\section{Summary and discussion}
In this paper, we proposed the simplest class of the inflationary $\alpha$-attractors in supergravity. The orthogonal and the cubic nilpotent conditions~\eqref{nilp}, \eqref{ort} not only reduce the number of physical degrees of freedom to one  real scalar, the inflaton, but also simplify the K\"ahler potential describing hyperbolic geometry~\eqref{KZ}, \eqref{KT}. As one can see from eqs. \rf{K}, \rf{KZ} for T-model and from eqs.~\rf{K1}, \rf{KT} for the E-model, the absence of the sinflaton scalar,  $Z-\bar Z=0$ and $T-\bar T=0$, keeps the K\"ahler metric to be the same as the one which follows from the  original more complicated \K\, potentials. These are  presented  in \rf{oldZ} and \rf{oldT}, respectively. The $\overline{D3}$ induced geometric inflation~\cite{McDonough:2016der,Kallosh:2017wnt} further enables us to build the models realizing inflation, supersymmetry breaking and a cosmological constant in the present universe. We applied this formalism to construct the simplest supersymmetric $\alpha$-attractors in Secs.~\ref{Tmodel} and \ref{Emodel}.

One can check that our inputs for the \K\, function in eqs. \rf{G}, \rf{Metric} for T-models and in \rf{GE}, \rf{MetricE} for E-models, lead to $\alpha$-attractor cosmological models with potentials in  eqs. \rf{V} for T-models and \rf{VE} for E-models. For this purpose  one can either check the derivation in appendix~\ref{appA}, or just run the corresponding Mathematica notebook. Either way, one can easily see that there is a relatively simple action with local supersymmetry, which results in data-compatible cosmological potentials for a single scalar field, inflaton.

\section*{Acknowledgements}
We are grateful to our collaborators on the related projects which helped us to find a  simple version of $\alpha$-attarctors, J. J. Carrasco, S. Ferrara, D. Freedman, A. Karlsson, A. Linde, J. Thaler, D. Roest, A. Van Proeyen, B. Vercnocke and T. Wrase.  RK and YY are supported by Stanford Institute for Theoretical Physics,  by the US National Science Foundation Grant  PHY-1720397 and by the Simons Foundation Origins of the Universe program (Modern Inflationary Cosmology collaboration).
\appendix
\section{Validation of new simple inflaton \K\, potentials}\label{appA}

Consider  the first inflaton shift symmetric K\"ahler potential for E-models~\cite{Carrasco:2015uma}, \cite{Kallosh:2017wnt}:
\begin{equation}
K=-\frac{3\alpha}{2}\log\left(\frac{(T+\bar T)^2}{4T\bar{T}}\right).\label{oldT}
\end{equation}
Obviously, our new K\"ahler potential in \rf{GE}, \rf{MetricE} for orthogonal inflaton superfield
\begin{equation}
K=-\frac{3\alpha}{2}\frac{(T-\bar{T})^2}{(T+\bar{T})^2}\label{newE}
\end{equation}
 is much simpler than \eqref{oldT}, for example, there are no $\log$'s.  We will show here that  \eqref{oldT} is actually reduced to \eqref{newE} when the cubic nilpotent condition~\eqref {nilportT},
$(T- \bar T)^k=0 \, ,k\geq 3$,
  is applied.
\begin{align}
K=&-\frac{3\alpha}{2}\log\left(\frac{(T+\bar T)^2}{4T\bar{T}}\right)&\nonumber\\
=&-\frac{3\alpha}{2}\log\left(\frac{(T+\bar T)^2}{(T+\bar T)^2-(T-\bar{T})^2}\right)\nonumber\\
=&\frac{3\alpha}{2}\log\left(1-\frac{(T-\bar{T})^2}{(T+\bar T)^2}\right)\nonumber\\
=&-\frac{3\alpha}{2}\frac{(T-\bar{T})^2}{(T+\bar T)^2}
\end{align}
where we have used $(T-\bar{T})^k=0, k\geq 3 $ in the last step. Thus, we have shown the equivalence between the old K\"ahler potential~\eqref{oldT} and the simplified one~\eqref{newE} for the orthogonal inflaton supermultiplet. 

In the disk variables $Z$ we have the first inflaton shift symmetric \K\,potential~\cite{Carrasco:2015uma},~\cite{Kallosh:2017wnt}
\begin{equation}
K=-\frac{3\alpha}{2}\log\left(\frac{(1-Z\bar Z)^2}{(1-Z^2)(1-\bar{Z}^2)}\right).
\label{oldZ}\end{equation}
 The $T$- and $Z$-variables are related by the Cayley transformations
\begin{equation}
T=\frac{1+Z}{1-Z},\qquad Z=\frac{T-1}{T+1}.
\end{equation}
It is known from  \cite{Carrasco:2015uma,Carrasco:2015rva}  that by making a change of variables $T=\frac{1+Z}{1-Z}$ one finds that \rf{oldT} becomes \rf{oldZ} and $(T-\bar{T})^k=0, k\geq 3$ condition becomes $(Z-\bar{Z})^k=0, k\geq 3$.
It follows that 
\begin{equation}
K=-\frac{3\alpha}{2}\log\left(\frac{(1-Z\bar Z)^2}{(1-Z^2)(1-\bar{Z}^2)}\right)\Big |_{(Z-\bar{Z})^k=0} = -\frac{3\alpha}{2}\frac{(Z-\bar{Z})^2}{(1-Z\bar{Z})^2}.
\label{TO}\end{equation}

\section{Details of Orthogonality Conditions in Local Supersymmetry}\label{appB}
Sinflaton $b$ depends on goldstino ${\Omega}^s$, on sgoldstino $s={\bar \Omega^s P_L \Omega^s\over 2 F^s}$, on $F^s$ and on the inflaton $\vp$ as follows:
\begin{align}
b= \frac{\rm i}{4}\Bigg[&\frac{\bar{\Omega}^s}{\bar{F}^s}\g^\mu P_L\frac{\Omega^s}{F^s}-\frac{\bar{s}}{\bar{F}^s}\left({\cal D}_\nu\frac{\bar{\Omega}^s}{F^s}\right)\g^\mu\g^\nu P_L \frac{\Omega^s}{F^s}\nonumber\\
&-\frac{s\bar{s}}{2|F^s|^2}\left({\cal D}_{\nu}\frac{\bar{\Omega}^s}{\bar{F}^s}\right)(\g^{\mu\nu\rho}+\g^\mu\eta^{\nu\rho}) P_L\left({\cal D}_{\rho}\frac{\Omega^s}{F^s}\right) -{\rm c.c.} \Bigg]{\cal D}_\mu\vp.
\label{b}
\end{align}
Inflatino depends on goldstino ${\Omega}^s$,  on $F^s$, on the sinflaton $b$, as shown above,  and on the inflaton $\vp$ as follows:
\be \label{inflatino} 
 P_R \Op = \left[\slashed{\D} (\vp-\rmi b)\right]P_L {\O^s\over F^s}.
\ee
Finally, the inflaton multiplet  auxiliary field  $F^\phi$ is  a  functional of goldstino ${\Omega}^s$,  sgoldstino $s={\bar \Omega^s P_L \Omega^s\over 2 F^s}$, of $F^s$, of the sinflaton $b$,  and on the inflaton $\vp$,
\be 
\bar{F}^\phi=\left({\cal D}_\nu \frac{\bar{\Omega}^s}{F^s}\right)P_L\g^\mu\g^\nu\frac{\Omega^s}{F^s}{\cal D}_\mu(\vp-{\rm i}b)+\frac{s}{F^s}{\cal D}^2(\vp-{\rm i}b).
\label{Fp}\ee
\bibliographystyle{JHEP}
\bibliography{lindekalloshrefs}
\end{document}